\documentclass[aps,onecolumn,groupedaddress,nofootinbib,amsmath,amssymb]{revtex4}
\usepackage{dsfont}
\usepackage{bm}
\usepackage{graphicx}
\begin{document}
\title[]{Thermodynamics of charged black holes with a nonlinear electrodynamics source}

\author{Hern\'an A. Gonz\'alez}\email{hdgonzal-at-uc.cl}
\affiliation{Departamento de F\'isica, Pontificia Universidad
Cat\'olica de Chile, Casilla 306, Santiago 22, Chile}
\author{Mokhtar Hassa\"{\i}ne}\email{hassaine-at-inst-mat.utalca.cl}
\affiliation{Instituto de Matem\'atica y F\'{\i}sica, Universidad de
Talca, Casilla 747, Talca, Chile}
\author{Cristi\'an Mart\'{\i}nez}\email{martinez-at-cecs.cl}
\affiliation{Centro de Estudios Cient\'{\i}ficos (CECS),
 Casilla 1469, Valdivia, Chile\\and
Centro de Ingenier\'{\i}a de la Innovaci\'on del CECS (CIN),
Valdivia, Chile}

%\preprint{{\tiny CECS-PHY-09/08} }

\begin{abstract} We study the thermodynamical properties of
electrically charged black hole solutions of a nonlinear
electrodynamics theory defined by a power $p$ of the Maxwell
invariant, which is coupled to Einstein gravity in four and higher
spacetime dimensions. Depending on the range of the parameter $p$,
these solutions present different asymptotic behaviors. We compute
the Euclidean action with the appropriate boundary term in the grand
canonical ensemble. The thermodynamical quantities are identified
and in particular, the mass and the charge are shown to be finite
for all classes of solutions. Interestingly, a generalized Smarr
formula is derived and it is shown that this latter encodes
perfectly the different asymptotic behaviors of the black hole
solutions. The local stability is analyzed by computing the heat
capacity and the electrical permittivity and we find that a set of
small black holes are locally stable. In contrast to the standard
Reissner-Nordstr\"om solution, there is a first-order phase
transition between a class of these non-linear charged black holes
and the Minkowski spacetime.

\end{abstract}

\maketitle

%%%%%%%%%%%%%%%%%%%%%%
\section{Introduction}
%%%%%%%%%%%%%%%%%%%%%%

The nonlinear electrodynamics models have been proved to be
excellent laboratories in order to circumvent some problems that
occur in the standard Maxwell theory. Indeed, the interest for
nonlinear electrodynamics has started with the precursor work of
Born and Infeld, whose main motivation was to modify the standard
Maxwell theory in order to eliminate the problem of infinite energy
of the electron \cite{Born:1934gh}. However, due to the fact that
the Born-Infeld model did not fulfill all the hope, nonlinear
electrodynamics theories become less popular. The recent renewal
interest on the nonlinear electrodynamics theories is essentially
due to their emergence in the context of low-energy limit of
heterotic string theory,  where a Gauss-Bonnet term coupled to
quartic contractions of Maxwell field strength appear, and black
hole solutions can be obtained \cite{HF}. It is also important to
mention that nonlinear electrodynamics theories are a powerful tool
for the construction of regular black hole solutions
\cite{regularBH}.

The thermodynamics properties of nonlinear electrodynamics theories
is also an active research area in the current literature, see e.g.
\cite{examplesTherm} in the case of the Born-Infeld model. A very
appealing property which is common to all the nonlinear
electrodynamic models lie in the fact that these theories satisfy
the zeroth and first law of black hole mechanics
\cite{Rasheed:1997ns}. This property renders more attractive the
studies of nonlinear electrodynamics models. However, in contrast
with the standard Maxwell theory, the formula of the total mass, the
so-called Smarr formula \cite{Smarr:1972kt}, does not hold {\it a
priori} for nonlinear electrodynamic theories \cite{Rasheed:1997ns}.
Indeed, in the case of the Einstein-Maxwell system, the Smarr
formula and the first law of black holes mechanics are closely
related and each of them can be derived from the other. This
correspondence between both formulas is due to the homogeneity of
the mass in terms of the area of the horizon and the electric charge
\cite{Gibbons:1976pt,Gauntlett:1998fz}. However, in the case of
nonlinear electrodynamic theories, this homogeneity property is not
longer present and there is {\it a priori} no good reason to obtain
a Smarr formula. Some attempts to generalize the Smarr formula in
order to be in accordance with the variations expressed by the first
law of black holes mechanics have shown to be possible for some
particular magnetic solutions in four dimensions
\cite{Breton:2004qa}. For completeness, we stress that the
derivation of the Smarr formula as well as of the first law of
thermodynamics are usually performed successfully for asymptotically
flat spacetime through the Komar integrals \cite{Bardeen:1973gs}.
However, their extension to rotating asymptotically anti-de Sitter
black hole solutions is not a simple task as it can be seen by
reading the current literature on the topic
\cite{Gibbons,Barnich:2004uw}.

In the present paper, we consider higher-dimensional gravity coupled
to a nonlinear electrodynamic source given by a rational power $p$
of the Maxwell invariant. This theory exhibits electrically charged
black hole solutions as long as the exponent $p$ belongs to the set
of rational number with odd denominator
\cite{Hassaine:2007py,Hassaine:2008pw}. Similar considerations and
motivations were previously analyzed in three dimensions in
\cite{Cataldo:2000we}. Recently, this nonlinear model has attracted
attention, and it has been used for obtaining Lovelock black holes
\cite{Hideki}, Ricci flat rotating black branes \cite{Hendi1},
magnetic strings \cite{Hendi2} and topological black holes in
Gauss-Bonnet gravity \cite{Hendi3}. A generalization of this model
for the non abelian case was considered in
\cite{Mazharimousavi:2009mb}. The black hole solutions derived in
\cite{Hassaine:2008pw} are divided in four ranges for the exponent
$p$, and each of them correspond to a different spacelike asymptotic
behavior of the metric. Among these different classes, there exist
black hole solutions with unusual asymptotic properties as those
that go asymptotically to the Minkowski spacetime slower than the
Schwarzschild spacetime. There also exists a range for $p$ for which
the solutions are not asymptotically flat and, their asymptotic
behavior is shown to grow slower that the Schwarzschild de Sitter
spacetime. Extremal black hole solutions generalizing the extremal
Reissner-Nordstr\"{o}m solution are also known for this model
\cite{Hassaine:2008pw}.

The main objective here is to provide a detailed analysis of the
thermodynamical properties of these black hole solutions. In
particular, we would like to explore the relation between the
thermodynamical properties and the different asymptotic behaviors of
the solutions. The thermodynamics of the electrically charged black
hole solutions will be performed through the Regge-Teitelboim
approach \cite{Regge:1974zd}, in which an appropriate boundary term
is added to the Euclidean action such that the total action presents
an extremum. On the other hand, owing to the fact that the Euclidean
action is related to the Gibbs free energy, the identification of
the mass and the charge, as well as the study of the global
stability of the black holes will be considerably facilitated. We
will derive a generalized Smarr formula by using the explicit
expressions of the temperature and the entropy. Interesting enough,
we will show that the different asymptotic behaviors of the black
hole solutions are reflected through this Smarr formula. In
addition, we will present two other different ways of obtaining of
the Smarr formula. The first derivation is operated through the
Komar integrals for all the class of solutions even those that are
not asymptotically flat. The same formula will also be obtained with
the use of a Noether conserved current which is associated to a
scale symmetry of the reduced action. The local stability of the
black holes are analyzed by computing the heat capacity and the
electrical permittivity. For a certain range of the exponent $p$,
small black hole solutions with positive and negative mass will be
shown to be locally stable. Finally, the global stability is studied
through the Gibbs free energy in order to determine whether the
electrically charged black hole solutions are most likely than the
Minkowski background. As a result, we will establish that there
exists a phase transition for a certain range of the exponent $p$.

The plan of the paper is organized as follows. In the next section,
we review the nonlinear electrodynamics model and its general black
hole solutions \cite{Hassaine:2008pw}. In Sec. III, the
thermodynamics properties of the system are studied through the
Euclidean Hamiltonian formalism. The mass $M$ and the charge $Q$ are
explicitly identified and a generalized Smarr formula is derived. In
Sec. IV the local and global stability of these black holes is
analyzed, while the last section is devoted to the conclusions and
further prospects. Finally, two appendices are devoted to the
different derivations of the Smarr formula.

%%%%%%%%%%%%%%%%%%%%%%%%%%%%%%%%%%%%%%%%%%%%%%%%%%%%%%%%%%%%%%
\section{Nonlinear electrodynamics and  black hole solutions}
%%%%%%%%%%%%%%%%%%%%%%%%%%%%%%%%%%%%%%%%%%%%%%%%%%%%%%%%%%%%%%
In \cite{Hassaine:2007py,Hassaine:2008pw}, a nonlinear
electrodynamics coupled to gravity in $d$ spacetime
dimensions\footnote{Along the present article we consider $d>3$.}
was considered. This model is described by the action
\begin{equation} \label{action}
I[g_{\mu \nu}, A_\mu]=\int d^{d}x\sqrt{-g}\left( \frac{R}{2\kappa
}-\alpha \left( F_{\mu \nu }F^{\mu \nu }\right) ^{p}\right),
\end{equation}
where $\kappa$ denotes the gravitational constant, $\alpha$ the
coupling constant for the electrodynamical action, $F_{\mu \nu}=
\partial_\mu A_\nu-\partial_\nu A_\mu$ is the field strength and $p$ is
a rational number whose range will be fixed later. The field
equations obtained by varying this action are given by
\begin{subequations}
\label{eqs}
\begin{eqnarray}
&&G_{\mu \nu }=4\kappa\alpha\left[p\,
F_{\mu\rho}F_{\nu}^{\,\,\,\rho}\,F^{p-1}-
\frac{1}{4}g_{\mu\nu}\,F^{p}\right], \label{EE} \\
&&\partial_{\mu}\left(\sqrt{-g}\,F^{\mu\nu}\,F^{p-1}\right)=0,\label{EE1}
\end{eqnarray}
\end{subequations}
where $F=F_{\alpha\beta}F^{\alpha\beta}$ is the Maxwell invariant.
The most general spherically symmetric solution with a radial
electric field was found in \cite{Hassaine:2007py} for the conformal
case $p=d/4$, and in \cite{Hassaine:2008pw} for $p$ belonging to the
set of rational numbers with odd denominators\footnote{This
restriction on $p$ arises from considering  spherically symmetric
real solutions with a purely electric radial field.}. The general
solution is described by the line element
\begin{equation} \label{1}
ds^{2}
=-N^{2}(r)f^{2}(r)dt^{2}+{\frac{dr^{2}}{f^{2}(r)}}+r^{2}d\Omega
_{d-2}^{2},
\end{equation}
with
\begin{subequations}
\label{generalsolution}
\begin{eqnarray}
&&N^{2}(r) =1 , \label{2} \\
&&f^{2}(r) =1-\frac{A}{r^{d-3}}+\frac{B}{r^{b}},  \label{f} \\
&&F_{tr} =\frac{C}{r^{\frac{d-2}{2p-1}}},  \label{F}
\end{eqnarray}
\end{subequations}
where for convenience we have defined
\begin{equation}
B=-\frac{2\kappa \alpha
(-1)^{p}C^{2p}2^{p}(2p-1)^{2}}{(d-2)(d-2p-1)},\qquad \quad
b=\frac{2(pd-4p+1)}{2p-1}. \label{b}
\end{equation}
These solutions contain black hole configurations depending on the
values of $d$ and $p$, and the integration constants $A$ and $C$.

Apart from the restriction of being a rational numbers with odd
denominators, the exponent $p$ can not belong to the set $(0,1/2]$
since in this case the scalar curvature of the solutions diverges at
the infinity. The form of the lapse function suggests a natural
partition depending on the exponent $b$ that appears in (\ref{f}).
These different ranges are given by $b>d-3$, $0<b<d-3$, $b<0$, and
$b=d-3$, which correspond to the different spacelike asymptotic
behaviors of the metric, as it is shown in Table I. Solutions of
type I have a behavior similar to the standard
Reissner-Nordstr\"{o}m one. By similar, we mean that  the charge
term in the metric decays faster than the mass term in the
asymptotic region. At the opposite, solutions of type II correspond
to black hole solutions which go asymptotically to the Minkowski
spacetime slower than the Schwarzschild spacetime. In the  case III,
the black hole solutions are not asymptotically flat, and their
asymptotic behavior is shown to grow slower that the Schwarzschild
de Sitter spacetime. Finally, the critical value $b=d-3$, which can
only occur for odd dimensions $d=2p+1$, corresponds to the limit
between the solutions which resemble to the standard
Reissner-Nordstr\"{o}m solution  (type I) and those for which the
asymptotic decaying to Minkowski spacetime are slower than the
Schwarzschild spacetime (type II). The spherically symmetric
solution in this special case involves a logarithmic dependance on
the radial coordinate and is given by
\begin{subequations}
\begin{eqnarray}
F_{tr} &=&\frac{C}{r},  \label{log} \\
f^{2}(r) &=&1-\frac{A}{r^{2p-2}}+\kappa \alpha
(-1)^{p}2^{p+1}C^{2p}\frac{ \ln r}{r^{2p-2}}. \label{flog}
\end{eqnarray}
\end{subequations}

\begin{table}
\begin{tabular}{|c|c|c|c|}
    \hline
Type &  $b$    &   $p$  & Remarks  \\
    \hline
I   & $b>d-3$& $1/2<p<(d-1)/2$ &Standard asymptotically flat case    \\
II  & $0<b<d-3$ &$p > (d-1)/2$ or $p< -1/(d-4)$ & Electric term with relaxed fall-off  \\
III  & $b\le 0$ &$-1/(d-4)\le p<0$ & Asymptotically non-flat case  \\
Log  &$b=d-3$ &$p= (d-1)/2$ with $d$ odd &Logarithmic case  \\
    \hline
\end{tabular}
\caption{Classification of the solutions depending on its asymptotic
behavior. The cases are labeled in the first column. In the second
column the different cases are defined by  the parameter $b$ and,
equivalently by $p$,  in the third column. In the last column we
describe  the main spacelike asymptotic feature for each case. }
\end{table}

%%%%%%%%%%%%%%%%%%%%%%%%%%%%
\section{Thermodynamics}
%%%%%%%%%%%%%%%%%%%%%%%%%%%%

It is well known that the partition function for a thermodynamical
ensemble can be identified with the Euclidean path integral in the
saddle point approximation around the Euclidean continuation of the
classical solution \cite{Gibbons:1976ue}.  In this case, the
Euclidean action $I_E$ evaluated on the classical solution is
related to the free energy $G$  of a thermodynamical ensemble by
$I_E= \beta G$, where $\beta$ is the inverse of the temperature
which corresponds to the period of the Euclidean time $\tau$.

As a  first step we write the action in the Hamiltonian form, and
since we are concerned only with the static, spherically symmetric
case without magnetic field,  it is enough to consider a
\textit{reduced} action principle. The class of the Euclidean metric
and electric potential to be considered are given by
$$
ds^2=N(r)^2f(r)^2d\tau^2+\frac{dr^2}{f(r)^2}+r^2d\Omega_{d-2}^2,
\qquad \quad \quad A= A(r) d\tau,
$$
where the radial coordinate $r$ belongs to $[r_+,\infty)$. In this
case, the Euclidean reduced action obtained from (\ref{action})
reads
\begin{eqnarray}
I_{E} =-\beta \;\Omega_{d-2}\int_{r_{+}}^{\infty }dr\left\{
\frac{(2p-1)\alpha
N(-2)^{\frac{p}{2p-1}}}{r^{\frac{d-2}{2p-1}}}\left( \frac{{\cal
P}}{4\alpha p} \right) ^{\frac{2p}{2p-1}} -\frac{d-2}{2\kappa
}Nr^{d-2}\left[ \frac{ (f^{2})^{\prime
}}{r}-\frac{d-3}{r^{2}}(1-f^{2})\right] +\phi {\cal P}^{\prime
}\right\} +K, \label{reducedaction}
\end{eqnarray}
where ${\cal P}\equiv 4\alpha p\,N_{\infty }\,r^{d-2}F^{p-1}F^{rt}$
is the rescaled canonical radial momentum,  $\phi \equiv A(r)$ is
the electrostatic potential,  $\Omega_{d-2}$ is the area of the
$d-2$-dimensional unit sphere. The boundary term $K$ appearing in
(\ref{reducedaction}) will be fixed by requiring that the action has
an extremum on-shell \cite{Regge:1974zd}. Moreover,  owing to the
fact that the Hamiltonian action is a linear combination of the
constraints, the value on the action on-shell is given by the
boundary term $K$.

The equations of motion obtained by varying the reduced action with
respect to $N$, $f^{2}$, ${\cal P}$ and $\phi $ are given by
\begin{subequations}
\label{redequations}
\begin{eqnarray}
&&\frac{(f^2)^{\prime }}{r}-\frac{d-3}{r^2}(1-f^2)= \frac{2\kappa
\alpha(2p-1)}{d-2}
\frac{(-2)^\frac{p}{2p-1}}{r^{\frac{2p(d-2)}{2p-1}}}\left(
\frac{{\cal P}}{4\alpha p}\right)^{\frac{2p}{2p-1}}, \\
&&N^{\prime }=0, \\
&&\phi^{\prime
}=\frac{N(-2)^{\frac{p}{2p-1}}}{2r^{\frac{d-2}{2p-1}}}\left(
\frac{{\cal P}}{4\alpha p}\right)^{\frac{1}{2p-1}}, \\
&&{\cal P}^{\prime }= 0.
\end{eqnarray}
\end{subequations}
Note that these equations are consistent with the original Einstein
equations (\ref{eqs}). The general solution reads
\begin{subequations}
\begin{eqnarray}
f^2(r) &=& \left\{
\begin{array}{lll}
\displaystyle
1-\frac{A}{r^{d-3}}+\frac{B}{r^{b}},  \qquad & \mbox{if} & b\neq d-3. \\
&& \\ 1-\frac{A}{r^{2p-2}}+\kappa \alpha (-1)^{p}2^{p+1}C^{2p}\frac{
\ln r}{r^{2p-2}}, \qquad & \mbox{if}& b=d-3,
\end{array}
\right.
\\
N(r) &=&N_{\infty} \\
\phi(r) &=& \left\{
\begin{array}{lll}
\displaystyle
\frac{2p-1}{d-2p-1}\frac{N_{\infty }C}{r^{\frac{d-2p-1}{2p-1}}}+\phi _{0},  \qquad & \mbox{if} & b\neq d-3. \\
&& \\ -N_{\infty }C \ln(r)+\phi _{0}, \qquad & \mbox{if}& b=d-3,
\end{array}
\right. \\
{\cal P}(r)&=&{\cal P}_{0}
\end{eqnarray}
\end{subequations}
where $N_{\infty }$, ${\cal P}_{0}$ , $A$ and $\phi _{0}$ are
integration constants (without loss of generality, we can set
$N_{\infty }=1$ and $\phi _{0}=0$), and  where we have defined
\begin{equation}
B=-\frac{2\kappa \alpha
(-2)^{\frac{p}{2p-1}}(2p-1)}{(d-2)(d-2p-1)}\left(\frac{{\cal
P}_{0}}{4 \alpha p}\right)^{\frac{2p}{2p-1 }}\quad \mbox{and} \quad
C=(-2)^{\frac{1-p}{2p-1}}\left({\cal P}_{0}\right)^{\frac{1}{2p-1 }}
({4\alpha p})^{\frac{1}{1-2p}}.
\end{equation}

In what follows, we consider the formalism of the grand canonical
ensemble, and hence we will consider the variation of the action
keeping fixed the temperature  $\beta^{-1}$ and the electric
potential, $\Phi=\phi(r_+)$. This variation gives bulk terms
proportional to the constraints, some surface terms and the
variation $\delta K$. As we shall see, the requirement that the
action has an extremum, i. e. $\delta I_{E}=0$ on-shell, will fix
properly the boundary term. The condition $\delta I_{E}=0$ implies
that the variation of the boundary term, which cancels out the extra
surface terms coming from variation of the action is given by
\begin{eqnarray}
\delta K =\beta \;\Omega_{d-2}\left[ -\frac{d-2}{2\kappa
}Nr^{d-3}\delta f^{2}+\phi \delta {\cal P}\right] _{r_{+}}^{\infty }
\equiv \delta K({\infty })-\delta K({r_{+}}), \label{current}
\end{eqnarray}
Since the metric solution $f^2(r)$ differs drastically for $b \neq
d-3$ and $b=d-3$, we shall consider both cases separately.
Interesting enough, we shall see that in both cases, the apparently
divergent contributions at the infinity will cancel yielding to a
finite and same expression in these two distinct cases.

For $b\neq d-3$, the variations of the fields at infinity
($r\rightarrow \infty$) are given by
\begin{eqnarray}
\delta f^{2}|_{\infty } &=&-r^{-(d-3)}\delta A+r^{-b}\delta B,
\label{F1}
\\
\delta {\cal P}|_{\infty } &=&\delta {\cal P}_{0},\label{F2}
\end{eqnarray}
and hence we have
\begin{equation}
\delta K({\infty})=\beta \;\Omega_{d-2}\left[\frac{d-2}{2\kappa
}\delta A+\left( - \frac{d-2}{2\kappa }\delta B+ \frac{2p-1}{d-2p-1}
C\delta {\cal P}_{0} \right) r^{d-3-b} \right]. \label{F3}
\end{equation}
For $b<d-3$, the contribution proportional to $r^{d-3-b}$ may blow
up at infinity, but since the factor between round brackets
multiplying this term identically vanishes, the variation of the
boundary term at infinity yields to a finite expression given by
\begin{eqnarray}
\delta K({\infty})=\frac{d-2}{2\kappa }\beta \;\Omega_{d-2} \delta
A. \label{kinfty}
\end{eqnarray}

For the case  $b=d-3$, which corresponds to an exponent $p=(d-1)/2$,
we have a similar derivation. Indeed, the variations of the fields
at infinity  read
\begin{eqnarray}
\delta f^{2}|_{\infty } &=&-r^{-(d-3)}\delta A +\kappa \alpha
(-1)^{p}2^{p+1}\delta C^{2p} r^{-(d-3)} \ln(r),  \label{L1}
\\
\delta {\cal P}|_{\infty } &=&\delta {\cal P}_{0},\label{L2}
\end{eqnarray}
and hence,
\begin{equation}
\delta K({\infty})=\beta \;\Omega_{d-2}\left[\frac{d-2}{2\kappa }A -
\Big((d-2) \alpha (-1)^{p}2^{p}\delta C^{2p}+  C\delta {\cal P}_{0}
\Big)\ln(r) \right].\label{L3}
\end{equation}
As in the previous case, since the following equality $ (d-2)\alpha
(-1)^{q}2^{q}\delta C^{2q}+ C\delta {\cal P}_{0}=0$ holds for
$p=(d-1)/2$, the logarithmic dependance disappears yielding  to the
same variation (\ref{kinfty}). We conclude that in both cases, the
integration of the variation (\ref{kinfty}) leads to a finite
expression given by
\begin{equation} \label{Kinf}
K({\infty})=\frac{d-2}{2\kappa }\beta \Omega_{d-2}A.
\end{equation}

In order to evaluate the variation of the metric function $f^2(r)$
at the horizon $r_{+}$, we use the fact that the solution satisfies
$f^2(r_{+})=0$ together with the condition that ensure the absence
of conical singularities at the horizon,  that is $(f^2)^{\prime
}|_{r_{+}}=4\pi /\beta $.  This condition fixes the temperature of
the ensemble while the electric potential $\phi$ and the variation
of ${\cal P}$ are evaluated directly,
\begin{eqnarray}
\delta f^2|_{r_{+}}&=&-(f^2)^{\prime
}|_{r_{+}}\delta{r_{+}}=-\frac{4\pi}{
\beta}\delta{r_{+}}, \\
\phi \,\delta{\cal P}|_{r_{+}}&=&\phi(r_{+})\delta {\cal P}_{0}.
\end{eqnarray}
The variation of the boundary term is easily integrated at the
horizon yielding
\begin{equation*}
K({r_{+}})=-\Phi \beta {\cal P}_{0}\Omega_{d-2}+\frac{2\pi }{\kappa
}\Omega_{d-2}r_{+}^{d-2}.
\end{equation*}
Finally, the on-shell Euclidean action, which reduces to the
boundary term $K=K({\infty })-K({r_{+}})$, reads\footnote{This holds
up to an arbitrary additive constant that is chosen to be equal to
zero by setting  $I_E=0$ in the case of flat spacetime.}
\begin{equation}
I_{E}=\beta \frac{d-2}{2\kappa }\Omega_{d-2}A-\beta \Phi {\cal
P}_{0}\Omega_{d-2}-\frac{2\pi }{\kappa }\Omega_{d-2}r_{+}^{d-2}.
\end{equation}
As it will be shown below, this expression is useful in order to
identify the thermodynamical quantities and in the study of the
global stability of the black hole solutions. Indeed, the Euclidean
action is related to the Gibbs free energy by $I_E= \beta G= \beta M
-\beta \Phi Q -S$. This relation allows to easily identify the mass
($M$), the electric charge ($Q$) and the entropy ($S$) as
\begin{subequations}
\label{thermoqtes}
\begin{eqnarray}
M &=&\left( \frac{\partial I_{E}}{\partial \beta }\right) _{\Phi
}-\frac{ \Phi }{\beta }\left( \frac{\partial I_{E}}{\partial \Phi
}\right) _{\beta }=
\frac{(d-2)\Omega_{d-2}}{2\kappa }A, \\
Q &=&-\frac{1}{\beta }\left( \frac{\partial I_{E}}{\partial \Phi
}\right)
_{\beta }=\Omega_{d-2}{\cal P}_{0}, \\
S &=&\beta \left( \frac{\partial I_{E}}{\partial \beta }\right)
_{\Phi }-I_{E}=\frac{2\pi }{\kappa }\Omega_{d-2}r_{+}^{d-2}.
\end{eqnarray}
\end{subequations}
As a first and direct consequence, these quantities must satisfy the
first law of thermodynamics $\delta M=T \delta S+\Phi \delta Q$
which is indeed the case. In a more unexpected way, these
expressions (\ref{thermoqtes}) permit to derive a generalized Smarr
formula. In order to achieve this task, we first express the event
horizon radius
 in terms of the thermodynamical quantities
(\ref{thermoqtes}) as
\begin{equation}
r_{+}^{d-3}=\frac{2\kappa M}{\Omega_{d-2}(d-2)}-\frac{\kappa
(2p-1)}{\Omega_{d-2}p(d-2) }Q\Phi. \label{outerh}
\end{equation}
Throughout this relation, the entropy and the temperature can be
written as
\begin{eqnarray*}
S &=&4\pi \left( \frac{2\kappa }{\Omega_{d-2}[2p\left( d-2\right)
]^{d-2}}\right) ^{\frac{1}{d-3}}\left[ 2pM-\left( 2p-1\right) Q\Phi
\right] ^{\frac{d-2}{d-3}
},\\
T &=&\frac{1}{4\pi }\left( \frac{p\Omega_{d-2}(d-2)}{\kappa }\right)
^{\frac{1}{ d-3}}\frac{2Mp(d-3)-2(pd-4p+1)Q\Phi }{\left(
2pM-(2p-1)Q\Phi \right) ^{\frac{ d-2}{d-3}}}.
\end{eqnarray*}
Remarkably enough, multiplying these two expressions and, after some
algebraic manipulations, we obtain a formula for the total mass
given by
\begin{equation}
M=\frac{d-2}{d-3}ST+\frac{pd-4p+1}{p\left( d-3\right) }Q\Phi,
\label{smarrf}
\end{equation}
which is nothing but a Smarr formula. We first observe that for
$p=1$, the relation (\ref{smarrf}) reduces to the well-known Smarr
formula for the $d-$dimensional Einstein-Maxwell black holes
\cite{Gauntlett:1998fz}. Another interesting value of the exponent
is given by $p=1/(4-d)$ for which the contribution of the charge in
the formula (\ref{smarrf}) disappears and the formula is similar to
the Smarr formula for the Schwarzschild solution. This is not
surprising since in this case, the charge contribution in the metric
(see (\ref{f}) and (\ref{b})) becomes constant and hence, the
solution resembles to the Schwarzschild metric but with a deficit
solid angle. In fact, this critical value of $p$ corresponds to the
transition between the solutions which asymptote the Minkowski
spacetime (type II) and those that go slowly than the Schwarzschild
de-Sitter solutions, i. e. $f^2(r) \sim r^{-b}$  with $0< -b < 2$
(type III). This peculiar transition is also reflected in the total
mass formula (\ref{smarrf}) by the fact that the term proportional
to the charge changes its sign. More precisely, the sign of the term
proportional to the charge in the case of solutions of type II is
opposite to the sign relative of those of type III, and the
transition is precisely operated at the critical value $p=1/(4-d)$.
Another value to consider is $p=0$ because of the apparent
singularity of the formula (\ref{smarrf}). However, a detailed
analysis shows that for $p=0$ the formula reduces to the standard
Smarr formula for the Schwarzschild-(anti) de Sitter
metric\footnote{The Smarr formula for AdS black holes and an
extended version of the first law including variations of $\Lambda$
have been studied in \cite{Kastor:2009wy}.} (with a cosmological
constant $\Lambda=\kappa\alpha$),
$$
M=\frac{d-2}{d-3}ST+\frac{2\alpha
\Omega_{d-2}}{(d-3)(d-1)}\left.r_{+}^{d-1}\right|_{p=0},
$$
where the expression of the horizon radius (\ref{outerh}) is
evaluated at $p=0$. This is not surprising since at the level of the
action, the value $p=0$ turns to be equivalent of having the
Einstein action with a cosmological constant. Within this
detailed study of the expression (\ref{smarrf}), we have pointed out
that the Smarr formula by its own reflects the different asymptotic
behaviors of the black hole solutions. Finally, we mention that the
Smarr law can also be derived in two different other manners: by
using the Komar integral even in the cases of the non asymptotic
flat solutions, (see Appendix A1) or by constructing a Noether
conserved current which is associated to a scaling symmetry of the
reduced action (\ref{reducedaction}), see Appendix A2.

For latter convenience, we now  express the mass and the temperature
in terms of the horizon $r_{+}$. Since we are working in the grand
canonical ensemble, these expressions must be written in terms of
the fixed state variables, namely the potential $\Phi$ and the
temperature $T$. A straightforward computation shows that the mass
$M$ expressed as a function of $r_+$ reads
\begin{eqnarray}
M=\frac{\Omega_{d-2}(d-2)}{2\kappa}\left[r_+^{d-3}+\frac{(2p-1)D}{d-2p-1}r_+^{d-2p-1}\right].
\end{eqnarray}
This relation is useful to analyze the behavior of the mass with
respect to the event horizon radius for the different classes of
solutions, cf. Fig. 1.
\begin{figure}
\centering
\includegraphics[scale=0.3]{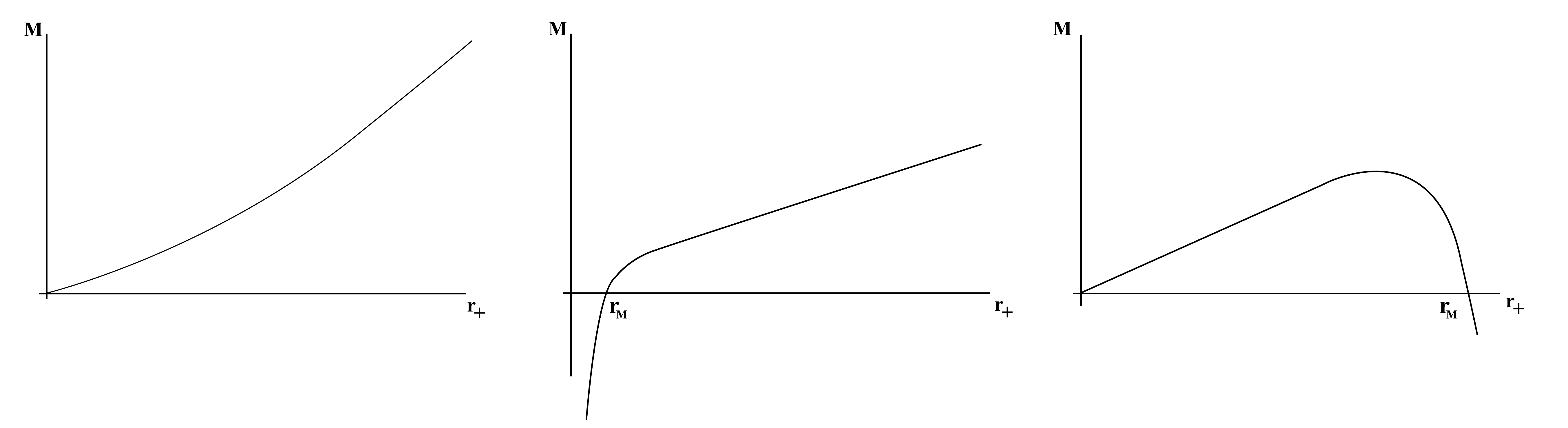}
\label{fig:Mass} \caption{Mass of the black hole in terms of the
event horizon radius $r_+$ at fixed electric potential $\Phi$. The
first graph corresponds to the solutions with $p \in
\left(\frac{1}{2},\frac{d-1}{2}\right)$, while the second one
represents the solutions with $p>\frac{d-1}{2}$. The last graph is
identified with the solutions with a negative exponent $p<0$. For
the two first graphs ($p>1/2$), the mass is a monotonous function of
the horizon radius, with the particularity that black hole solutions
with negative mass are only allowed for $p>(d-1)/2$. This fact
occurs when $r_+ <r_{\rm M}$, where $r_{\rm M}$ corresponds to the
radius for which the mass vanishes. In contrast, for $p<0$ the mass
is not a monotonous function and has a local maximum, as it is shown
in the last graph. Moreover, in the range ${1}/(4-d)<p<0$, only
black hole solutions with positive mass can be exhibited, since for
this interval of $p$ the negative mass solutions have a negative
temperature (see Fig. 2). For the remaining range $p<1/(4-d)$, the
mass can be negative since its corresponding temperature is
positive.}
\end{figure}
On the other hand, the square temperature is given by
\begin{eqnarray*}
T^2 &=&\left[\frac{1}{4\pi}\left.\frac{df^{2}}{dr}\right|_{r+}\right]^2,\\
\end{eqnarray*}
which, after some algebraic manipulations, can be rewritten as
\begin{equation} \label{temp-phi}
T^2=\left[\frac{1}{4\pi }\left(
\frac{d-3}{r_{+}}-\frac{D}{r_{+}^{2p-1}}\right)\right]^2,
\end{equation}
and where the constant $D$ is given by
\begin{equation*}
D=-2\kappa \alpha (-2)^{p}(d-2)^{-1}(2p-1)^{1-2p}(d-2p-1)^{2p}\Phi
^{2p}.
\end{equation*}
It is important to mention that, due to the positive energy
condition \cite{Hassaine:2008pw}, the constant $D$ is positive for
all the values of the exponent $p$. For clarity, in Fig. 2, we
sketch the temperature as a function of the horizon radius for the
different range of the parameter $p$.

\begin{figure}
\centering
\includegraphics[scale=0.4]{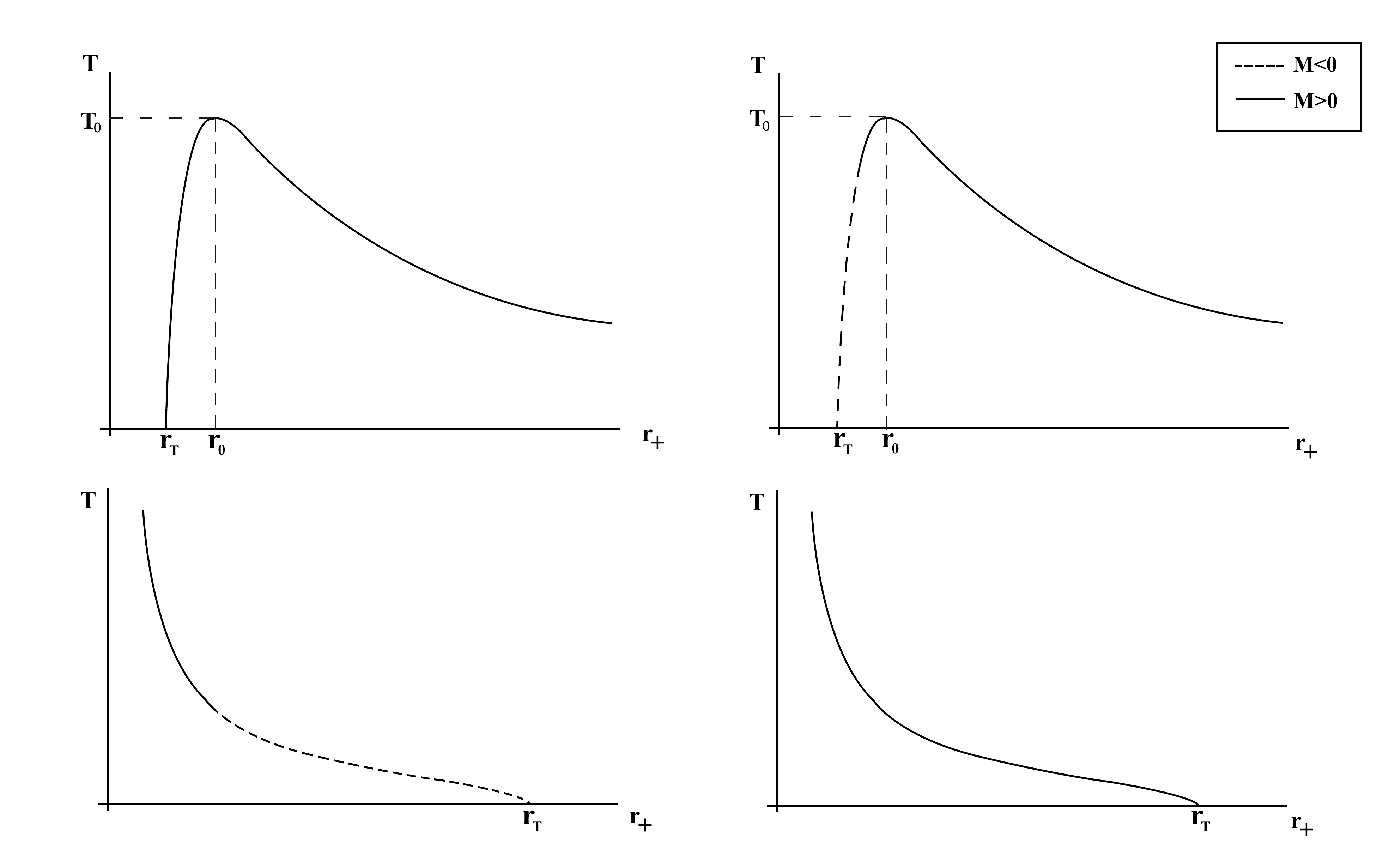}
\caption{Temperature against the horizon $r_+$ for different ranges
of the exponent $p$  at fixed electric potential $\Phi$. At the top
left, the graph corresponds to the range
$p\in(\frac{1}{2},\frac{d-1}{2})$, while the top right graph
represents the solutions with $p>\frac{d-1}{2}$. The last graphs are
respectively for $p<\frac{1}{4-d}$ and $p\in (\frac{1}{4-d},0)$. For
the exponent $p$ ranging over the set $(\frac{1}{2},\frac{d-1}{2})$,
the asymptotic behavior of the solution is similar to the
Reissner-Nordstr\"om black hole, and the analogy is also valid
concerning the thermodynamics. In particular, for large value of the
horizon, the temperature goes to zero while it reaches its maximum
$T_0$ for a finite value of the horizon radius $r_0$. In addition,
there exists a value $r_{\rm T}$ for which the temperature vanishes,
which corresponds to an extremal black hole. For $p>(d-1)/2$, the
main difference with the previous case lies in the fact that there
exist black holes with positive and negative mass, and the extremal
solution is possible only for negative mass. For $p<1/(4-d)$, the
temperature behavior is quite different since it is a decreasing
function of the event horizon,  and as before, the extremal solution
is possible only with a negative mass. Finally, for $1/(d-4)<p<0$,
the temperature is similar to the previous case, with the exceptions
that the mass is always positive, and the value for which the
temperature vanishes does not correspond to an extremal black hole.
} \label{fig:T111}
\end{figure}

%%%%%%%%%%%%%%%%%%%%%%%%%%%%%%%%%%%%
\section{Local and global thermodynamic stability}
%%%%%%%%%%%%%%%%%%%%%%%%%%%%%%%%%%%%%%%
The thermodynamic stability of a system can be considered from many
different point of views depending on which thermodynamical
variables or state functions one is considering. Usually, in order
to study the stability of a system, it is common to consider small
fluctuations of the state functions around the equilibrium, and
since the first order terms vanish, the stability is only a
statement about the second order variations. An equivalent manner of
studying the local stability can be done by analyzing  the sign of
the heat capacity  $C_{\Phi}$ at constant potential
\begin{eqnarray}
C_{\Phi} \equiv T\left( \frac{\partial S}{\partial T}\right) _{\Phi
}, \label{cp}
\end{eqnarray}
as well as, the sign of the electrical permittivity $\epsilon_T$ at
constant temperature
\begin{eqnarray}
\epsilon_T \equiv\left(\frac{\partial Q}{\partial\Phi}\right)_T.
\label{et}
\end{eqnarray}
From these definitions, it is clear that the heat capacity (resp.
the electrical permittivity) gives information about the thermal
stability with respect to the temperature fluctuations (resp. to the
electrical fluctuations). The positivity of the heat capacity,
$C_{\Phi}\geq 0$, is a necessary condition for the system to be
locally stable and, this is a direct consequence of the definition
(\ref{cp}) and the fact that the entropy is proportional to the size
of the black hole \cite{Chamblin:1999hg}. On the other hand, the
black holes will be electrically unstable under electrical
fluctuations if the electrical permittivity is negative. The
physical interpretation of this instability is explained by the fact
that in this case the potential decreases while the system acquires
more charge, and hence the system may leave easily the equilibrium
state \cite{Chamblin:1999hg}. The local stability of a system will
be ensured only if there exists a range of the horizon radius for
which the quantities $C_{\Phi}$ and $\epsilon_T$ are both positives.

In order to express the heat capacity (\ref{cp}) as well as the
electrical permittivity (\ref{et}) in terms of the horizon radius,
we rewrite them as
\begin{eqnarray}
C_{\Phi} =T\left(\frac{\partial T}{\partial r_{+}}\right)^{-1}
_{\Phi }\left( \frac{\partial S}{\partial r_{+}}\right) _{\Phi }
=-\frac{2\pi }{\kappa }\Omega_{d-2}(d-2)r_{+}^{d-2}\left( \frac{%
D-(d-3)r_{+}^{2p-2}}{(2p-1)D-(d-3)r_{+}^{2p-2}}\right), \label{Cphi}
\end{eqnarray}
and
$$
\epsilon_T =\left(\frac{\partial \Phi }{\partial r_{+}}\right)
_{T}^{-1} \left( \frac{\partial Q}{\partial r_{+}}\right) _{T}.
$$
However, since we have only the dependence of the charge $Q$ in
terms of the potential at constant horizon, we have to rewrite each
factor of the above formula as
\begin{eqnarray*}
&&\left( \frac{\partial Q}{\partial r_{+}}\right) _{T} =-\left( \frac{%
\partial T}{\partial Q}\right) _{r_{+}}^{-1}\left( \frac{\partial T}{%
\partial r_{+}}\right) _{Q}, \\
&&\left( \frac{\partial \Phi }{\partial r_{+}}\right) _{T}=-\left( \frac{%
\partial T}{\partial \Phi }\right) _{r_{+}}^{-1}\left( \frac{\partial T}{%
\partial r_{+}}\right) _{\Phi }.
\end{eqnarray*}
Combining these last expressions, the permittivity is found to be
\begin{eqnarray}
\epsilon_T &=&\left( \frac{\partial Q}{\partial \Phi }\right) _{r_{+}}\left( \frac{%
\partial T}{\partial r_{+}}\right) _{\Phi }^{-1}\left( \frac{\partial T}{%
\partial r_{+}}\right) _{Q}\nonumber\\
&=&-\frac{p(d-2)D}{\kappa\Phi^2(d-2p-1)}\Omega_{d-2}r_+^{d-2p-1}
\left(\frac{(2pd-6p+1)D-(2p-1)(d-3)r_+^{2p-2}}{(2p-1)D-(d-3)r_+^{2p-2}}\right).
\label{epsilonT}
\end{eqnarray}
From the above formulas (\ref{Cphi}) and (\ref{epsilonT}), it is
clear that there is a strong influence of the exponent $p$ on the
sign of these two expressions. Moreover, for a positive exponent
$p$,  both expressions will diverge at certain value of horizon
radius $r_0$, reflecting a radical change in the thermodynamical
local stability of the system. Indeed,  the sign of the specific
heat and electrical permittivity is flipped at $r_0$. For these
reasons, it is interesting to plot $C_{\Phi}$ as well as
$\epsilon_T$ against the event horizon $r_+$ for the different
ranges of the exponent $p$.

\begin{figure}
\centering
\includegraphics[scale=0.32]{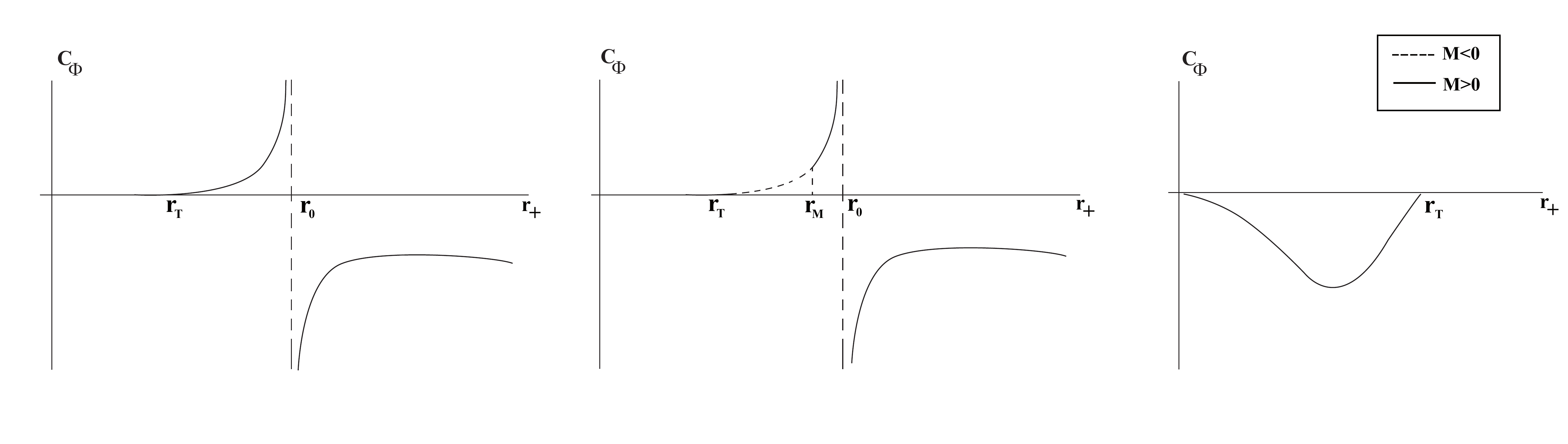}
\label{fig:Heat} \caption{Specific heat at fixed potential with
$p\in(\frac{1}{2},\frac{d-1}{2})$, $p>\frac{d-1}{2}$ and $p<0$.}
\end{figure}

\begin{figure}
    \centering
        \includegraphics[scale=0.35]{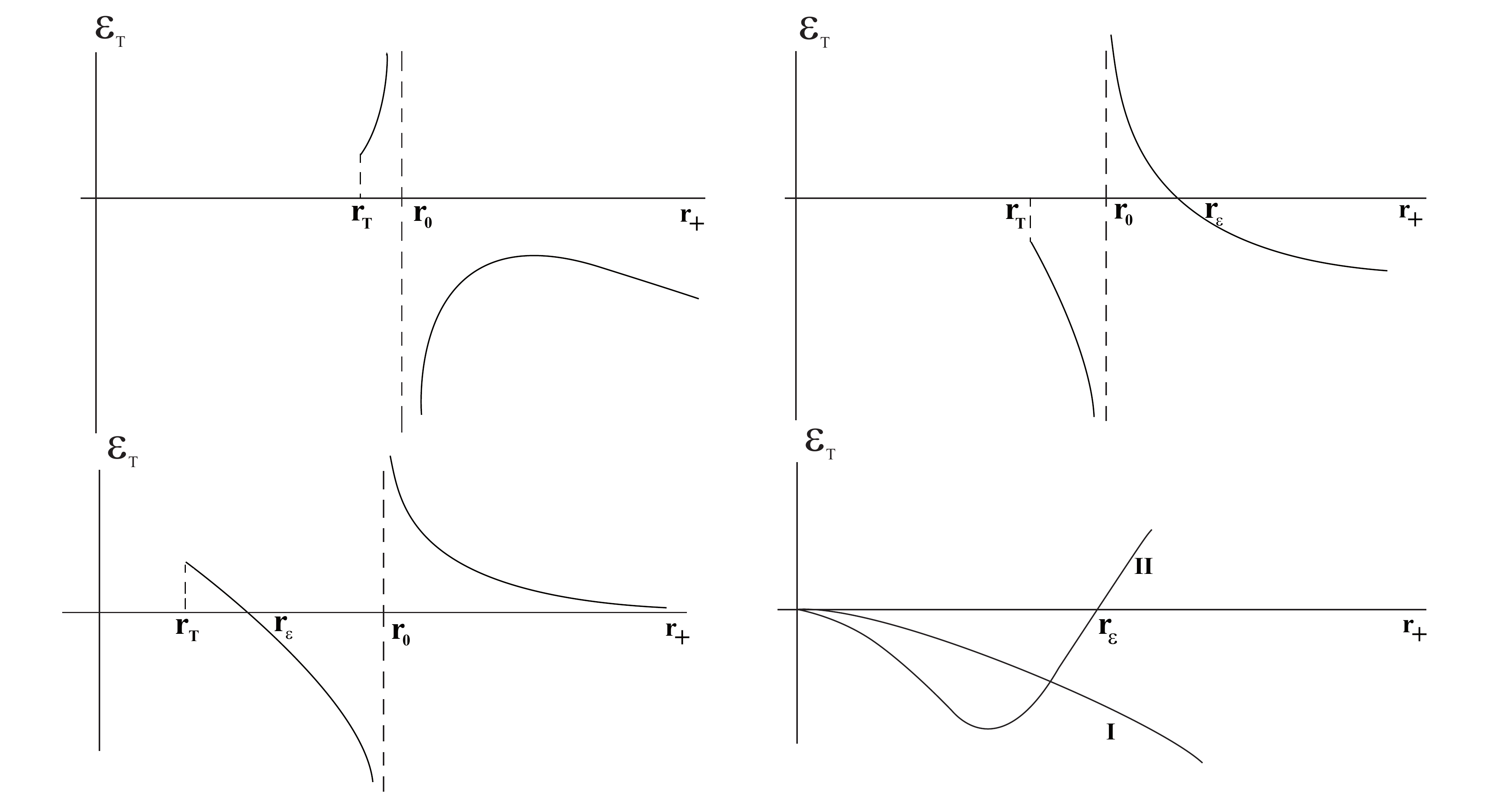}
    \caption{ The electrical permittivity against $r_+$ for $p\in(\frac{1}{2},1)$,
    $p\in(1,\frac{d-1}{2})$ and $p>\frac{d-1}{2}$. For  $p<0$ , there are two branches: the branch I
    corresponds to
$p>\frac{1}{2(3-d)}$ while the branch II is relative to
$p<\frac{1}{2(3-d)}$.}
    \label{fig:T11}
\end{figure}
For the range $p\in (1/2, (d-1)/2)$, there exist only black holes
with positive mass. In this case, the heat capacity (cf. Fig. 3) for
$p\not=1$ have a positive and negative branches and present a
vertical asymptote at $r_0$. For black hole with horizon $r_+\in
(r_T,r_0)$, the heat capacity is positive, and hence the stability
of the black hole is ensured for the thermal fluctuations. At the
Reissner-Nordstr\"{o}m limit, i.e. $p=1$, the singularity disappears
and the heat capacity is always negative reflecting the instability
of the solution. For the electrical permittivity (cf. Fig. 4), the
analysis is divided in two parts. For $p\in(1/2,1)$ and for the
region $r_+\in (r_T, r_0)$, the permittivity is positive and hence
we conclude that the system is locally stable. In contrast, for
$p\in(1,(d-1)/2)$ the system is unstable since there are no common
regions where the heat capacity and the permittivity are both
positive. For $p>(d-1)/2$, only small black hole with $r_+\in (r_T,
r_{\epsilon})$ and with negative mass are locally stable. Finally
for $p<0$, independently of the sign of the mass, the solution is
always unstable since the heat capacity is negative and hence the
system is locally unstable. The main result is that the nonlinear
electrodynamics theory considered in this paper permits the
existence of locally stable black hole solutions. This result
emphasizes the importance plays by the nonlinearity and is put in
opposition with the local thermal instability of the standard
Reissner-Nordstr\"{o}m black hole.

We now turn to the study of the global stability in order to
determine whether our solutions are thermodynamically preferred over
the Minkowski background.  The Gibbs free energy $G=I_E/\beta$ is an
appropriate state function to compare configurations in the grand
canonical ensemble. For example, it is well-known that in the
standard Einstein-Maxwell theory, the Minkowski spacetime is always
favored over the Reissner-Nordstr\"{o}m black hole since in this
case the free energy of this latter is positive.
\begin{figure}
\centering
\includegraphics[scale=0.4]{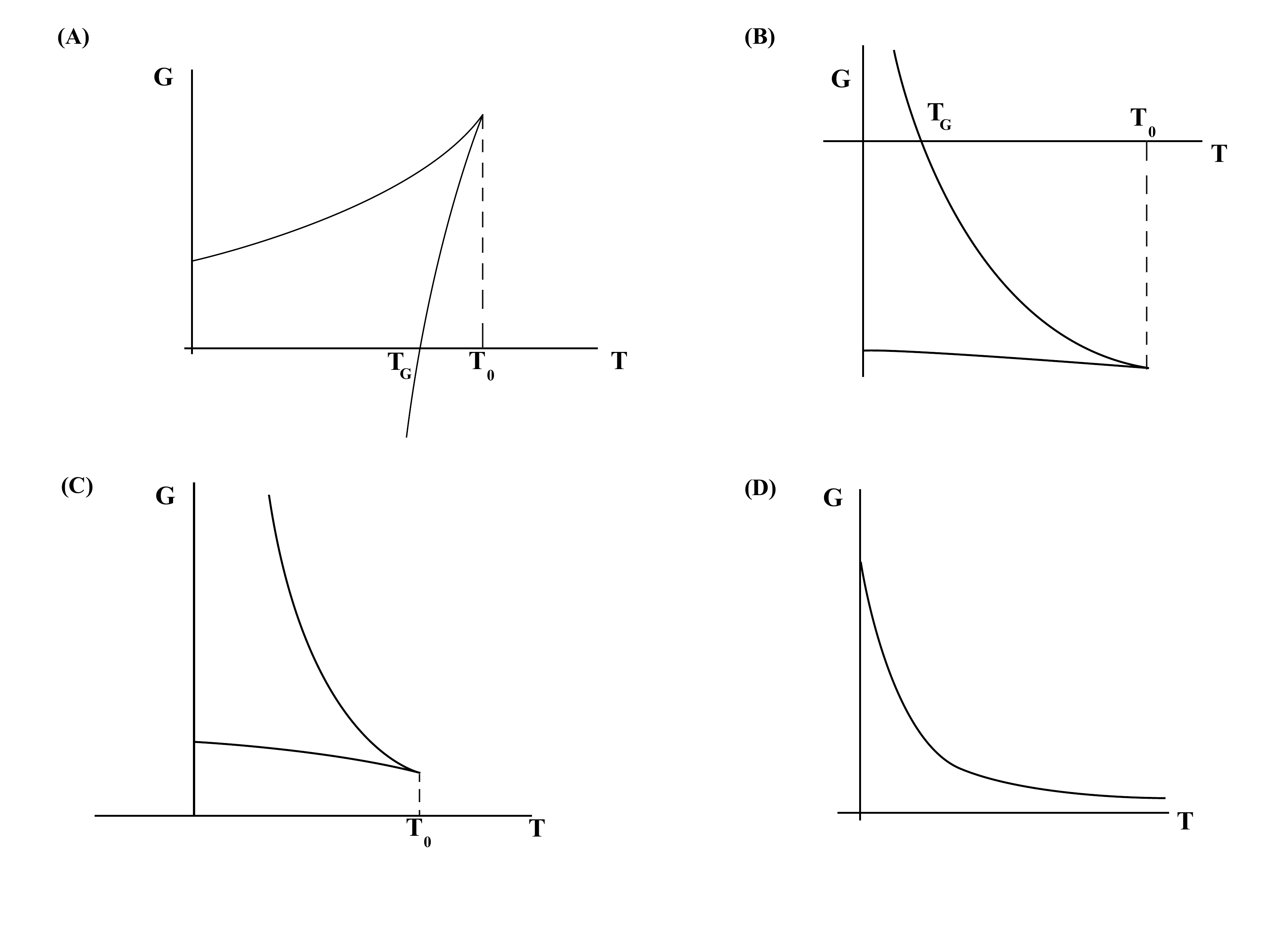}
\caption{The Gibbs free energy in terms of temperature for
$p\in(\frac{1}{2},1)$, $p\in(1,\frac{d-1}{2})$, $p>\frac{d-1}{2}$
and $p<0$ at fixed electric potential $\Phi$. A first-order phase
transition can be observed only in the first graph at the
temperature $T_G$.}
    \label{fig:T1}
\end{figure}
In our case (cf. Fig 5) and for $p<0$, since the temperature is a
monotonous function of the event horizon radius, the free energy is
a positive decreasing function and hence the Minkowski background is
more likely than the black hole configurations. For $p>(d-1)/2$ and
for a fixed temperature two black hole configurations with different
free energy and size coexist. In this case, in spite of the fact
that both solutions are less likely than the Minkowski background,
there is a non-vanishing probability that the black hole with the
larger $r_+$ decays into the smaller one. In the case where the
parameter $p\in(1/2,1)$ and for the temperature $T<T_G$, the larger
black holes are thermodynamically preferred over the Minkowski
background, and since for $T> T_G$ the black hole branch has a
larger free energy than the Minkowski spacetime a first-order phase
transition occurs at $T=T_G$. The situation is drastically different
for $p\in(1,\frac{d-1}{2})$. Indeed, in this case the most probable
configuration is that concerned with the small black hole branch,
but in this case, phase transitions are not observed.

%%%%%%%%%%%%%%%%%%%%%%%%%%%%%%%%%%%%%%%
\section{Conclusions and comments}
%%%%%%%%%%%%%%%%%%%%%%%%%%%%%%%%%%%%%%%
We have studied the thermodynamical properties of  black hole
arising as solutions of higher-dimensional gravity coupled to a
nonlinear electrodynamics theory given as a power $p$ of the Maxwell
invariant. These solutions have the peculiarity of having different
asymptotic behaviors depending mainly on the range of the exponent
$p$, including non asymptotically flat spacetimes. Metrics with
asymptotic relaxed fall-off has attracted much attention
\cite{slowfall} for AdS gravity coupled to a scalar field with mass
at or slightly above the Breitenlohner-Freedman bound \cite{BF}.
This theory allows a large class of asymptotically AdS spacetimes
where the charges can be properly defined. In the present work we
have identified the integrations constants with the mass and the
electric charge by using the Euclidean action in its Hamiltonian
form. We have adopted the grand canonical ensemble by keeping fixed
the temperature and the electric potential. We have shown that
although the variation of the dynamical fields may diverge at the
infinity, these divergences are canceled yielding a finite Euclidean
action. From this regularized action we derive the different
thermodynamical quantities and a generalized version of the Smarr
formula is obtained. The different asymptotic behaviors of the
solutions have been shown to be encoded by the Smarr formula. We
have also proposed a derivation of the Smarr formula by using a
Noether conserved current which results of the scale invariance of
the reduced action. A similar derivation of the Smarr formula  has
been obtained in the case of scalar hairy black holes coupled
minimally to the three-dimensional Einstein gravity
\cite{Banados:2005hm}. Note that in this reference the matter field
vanishes asymptotically,  but  this condition is not fulfill in our
case  since the Maxwell potential may be divergent as $r$ goes to
infinity depending on the range of the exponent $p$. However, these
divergences, which also appear in the Komar integrals, are canceled
and a Smarr formula can be written.

The local thermodynamic stability of the solutions have been
analyzed through the heat capacity and the electrical permittivity.
We have shown that, contrarily to the Einstein-Maxwell solution,
there exist small black holes  that are locally stable in the sense
that there exists a range for the horizon radius for which the heat
capacity and the electrical permittivity are both positive. An
interesting problem to be dealt is the study of the classical
dynamic stability of the model considered here and its possible
relation with the local thermodynamic stability.

It is worth to be noted that there exists some ranges of the
exponent $p$ for which the black hole solutions are preferred over
the Minkowski background, and a first-order phase transition appears
in the case $p\in (1/2,1)$. This situation is clearly in contrast
with what occurs in the standard Einstein-Maxwell case where the
flat spacetime is always likely than the Reissner-Nordstr\"om
solution.

%%%%%%%%%%%%%%%%%%%%%%%%%%%%%%%%%%%%%%%%%%%%%
\appendix
\section{Other derivations of the Smarr formula}
%%%%%%%%%%%%%%%%%%%%%%%%%%%%%%%%%%%%%%%%%%%%%%%%%%

%%%%%%%%%%%%%%%%%%%%%%%%%%%%%%%%%%%%%%%%%%%%%%%%
\subsection{Smarr formula from Komar integral}
%%%%%%%%%%%%%%%%%%%%%%%%%%%%%%%%%%%%%%%%%%%%%%%%
Owing to the fact that the black hole solution (\ref{1}) is time
translation invariant, the vector field $\xi =\partial _{t}$ is a
Killing vector field. We are going to evaluate the Komar integral
for this vector over the $(d-2)$-sphere at spatial infinity with
induced metric $\gamma$,
\begin{subequations}
\begin{eqnarray}
\label{de1}\oint_{\infty}dS_{\mu \nu }\nabla ^{\mu }\xi ^{\nu }
&=&-\oint_{\infty}d\Omega _{d-2}\sqrt{\gamma }\
r^{d-2}g^{rr}g^{tt}g_{tt,r}\\
 \label{de2}&=&-2\kappa \frac{d-3}{d-2}M+
2\kappa \frac{pd-4p+1}{p\left( d-2\right) }\left[Q\phi(r)
\right]_\infty.
\end{eqnarray}
\end{subequations}
Using the Stokes theorem and the properties of the Killing vector,
the expression (\ref{de1}) can be rewritten as
\begin{eqnarray}
\oint_{\infty}dS_{\mu \nu }\nabla ^{\mu }\xi ^{\nu } &=&\int_{{\cal
H}}dS_{\mu \nu }\nabla ^{\mu }\xi ^{\nu }+2\int_{\Sigma }dS_{\mu
}R_{\sigma }^{\mu }\xi ^{\sigma },
\end{eqnarray}
where $\Sigma$ is a spacelike hypersurface covering the region
between the outer horizon ${\cal H}$ and the spatial infinity. Using
the equations of motion, this last expression becomes
\begin{eqnarray}
\label{e3} \oint_{\infty}dS_{\mu \nu }\nabla ^{\mu }\xi ^{\nu }
=-2\kappa TS-2\kappa \left( \frac{pd-4p+1}{\left( d-2\right)
p}\right) Q\left[\phi(r_+) -\phi (r)\right]_\infty.
\end{eqnarray}
Equating the expressions (\ref{de2}) and (\ref{e3}), the terms
proportional to the electric potential at infinity are canceled out
and one recovers the Smarr formula (\ref{smarrf}).

%%%%%%%%%%%%%%%%%%%%%%%%%%%%%%%%%%%%%%%%%%%%%
\subsection{Smarr formula from Noether conserved current}
%%%%%%%%%%%%%%%%%%%%%%%%%%%%%%%%%%%%%%%%%%%%%%%%%%
The Smarr formula (\ref{smarrf}) can also be obtained from a Noether
current density by observing that the reduced action
(\ref{reducedaction}), or equivalently the field equations
(\ref{redequations}) are invariant under the following scaling
transformations
\begin{eqnarray*}
\bar{r}=\sigma\,r,\qquad \bar{{\cal
P}}(\bar{r})=\sigma^{\frac{pd-4p+1}{p}}{\cal P}({r}),\qquad
\bar{\phi}(\bar{r})=\sigma^{\frac{-pd+4p-1}{p}}{\phi}({r}),\qquad
\bar{f}^2(\bar{r})=f^2(r),\qquad \bar{N}(\bar{r})=\sigma^{3-d}N(r),
\end{eqnarray*}
where $\sigma$ is the parameter associated to the scale symmetry.
Note that a similar scale symmetry has been observed in the case of
three-dimensional scalar field minimally coupled to gravity
\cite{Banados:2005hm}. A straightforward application of the Noether
theorem yields the following current
\begin{eqnarray}
C(r)=\phi\left[r{\cal P}^{\prime}-\frac{pd-4p+1}{p}{\cal
P}\right]-\frac{d-2}{2\kappa}(f^2)^{\prime}Nr^{d-2},
\end{eqnarray}
which is conserved, i.e. $\partial_r C(r)=0$. This conservation law
can also be proved directly using the equations of motion
(\ref{redequations}). Evaluating this expression at infinity and at
the horizon $r_+$, one gets
$$
C(\infty)=-\frac{d-3}{\Omega_{d-2}}M,\qquad\qquad
C(r_+)=-\frac{1}{\Omega_{d-2}}\left[ST(d-2)+\frac{pd-4p+1}{p}Q\Phi\right].
$$
Finally, using the fact that $C$ is a constant, $C(\infty)=C(r_+)$,
the Smarr formula (\ref{smarrf}) is recovered.

\acknowledgments We thank Fabrizio Canfora, Julio Oliva and Ricardo
Troncoso for useful discussions. This work has been partially
supported by grants 1061291, 1071125, 1085322, 1095098 and 1090368
from FONDECYT and by the project Redes de Anillos R04 from CONICYT.
The Centro de Estudios Cient\'{\i}ficos (CECS) is funded by the
Chilean Government through the Millennium Science Initiative and the
Centers of Excellence Base Financing Program of Conicyt. CECS is
also supported by a group of private companies which at present
includes Antofagasta Minerals, Arauco, Empresas CMPC, Indura,
Naviera Ultragas and Telef\'onica del Sur. CIN is funded by Conicyt
and the Gobierno Regional de Los R\'{\i}os.

%%%%%%%%%%%%%%%%%%%%%%%%%%%
%%%%%%%%%%%%%%%%%%%%%%%%%%%

\end{document}